\documentclass[final,3p,times]{elsarticle}
\usepackage{graphicx}
\usepackage{subfigure}
\usepackage{amssymb}
 \usepackage{lineno}
\usepackage{amsmath}
\usepackage{mathrsfs}
\usepackage{dcolumn}
\usepackage{psfrag}
\usepackage{comment}
\usepackage{color}

\newcommand{\beq}{\begin{equation}}
\newcommand{\eeq}{  \end{equation}}
\newcommand{\beqa}{\begin{eqnarray}}
\newcommand{\eeqa}{  \end{eqnarray}}

\def\XXint#1#2#3{{\setbox0=\hbox{$#1{#2#3}{\int}$}
     \vcenter{\hbox{$#2#3$}}\kern-.5\wd0}}

\definecolor{red}{rgb}{1,0,0}
\definecolor{blue}{rgb}{0,0,1}

\def\strutdepth{\dp\strutbox}
\def\nw#1{\strut\vadjust{\kern-\strutdepth\vtop to0pt{\vss\hbox to\hsize
{\hskip\hsize\hskip5pt$\leftarrow$\hss\strut}}}{\em #1}}

\journal{Journal of Computational Physics}

\begin{document}

\begin{frontmatter}
\title{A general method to remove the stiffness of PDEs}


\author[LD]{Laurent Duchemin}
\ead{duchemin@irphe.univ-mrs.fr}
\author[JE]{Jens Eggers}
\ead{Jens.Eggers@bristol.ac.uk}

\address[LD]{Aix Marseille Universit\'e, CNRS, Centrale Marseille, IRPHE UMR 7342, F-13384, Marseille, France}
\address[JE]{Department of Mathematics - University of Bristol, 
University Walk, Bristol BS8 1TW, United Kingdom}


\begin{abstract}
A new method to remove the stiffness of partial differential equations 
is presented. Two terms are added to the right-hand-side of the PDE~: 
the first is a damping term and is treated implicitly, the second 
is of the opposite sign and is treated explicitly. A criterion for 
absolute stability is found and the scheme is shown to be convergent. 
The method is applied with success to the mean curvature flow equation, 
the Kuramoto--Sivashinsky equation, and to the Rayleigh--Taylor 
instability in a Hele-Shaw cell, including the effect of surface tension. 
\end{abstract}

\begin{keyword}
Stiff set of partial differential equations \sep Kuramoto--Sivashinsky 
\sep Hele-Shaw \sep Birkhoff--Rott integral \sep surface tension
\end{keyword}

\end{frontmatter}




\section{Introduction}

Many partial differential equations (PDEs) which arise in physics 
or engineering involve the computation of higher-order spatial derivatives.
These higher-order derivatives may have several origins, most 
commonly diffusion, where the time derivative of the variable 
is determined by 2nd-order spatial derivatives on the right-hand-side
(RHS) of the equation.
In the physics of interfaces, surface tension is often taken into account 
through Laplace's law, which introduces second or third-order spatial
derivatives. If diffusion is driven by surface tension, derivatives can 
easily be of fourth order, for example in surface diffusion \cite{Mull57}. 

If one advances the solution using an explicit integration scheme
(RHS evaluated at the old time step), and the order of the highest 
derivative is $m$, then for the method to be stable, the time step 
$\delta t$ is required to scale like $\delta x^m$, where $\delta x$ is 
the grid spacing:
\beq
\delta t = C \delta x^m.
\label{stab_explicit}
\eeq
The constraint (\ref{stab_explicit}) on the time step is known as 
numerical {\em stiffness}. It corresponds to the decay time of the 
fastest modes present in the system, excited on the scale of the 
numerical grid. In practice, however, the physical interest lies 
in describing features on a scale much larger than $\delta x$. Thus,
in particular if $m=2$ or higher, (\ref{stab_explicit}) imposes a 
time step much smaller than warranted by the physical time scale of 
interest, and renders explicit schemes impractical. 

A way of removing (\ref{stab_explicit}) as a constraint on the time step 
is to use an implicit scheme, for which the RHS is evaluated at the 
yet-to-be-computed time. In general, this involves the solution of a
(nonlinear) set of equations to compute the solution at the new time step.
In the case of the linear diffusion equation with constant coefficients, 
this can be done very efficiently. If the transport coefficients vary in 
space, or depend on the solution (quasilinear case), the method becomes
more cumbersome, and usually requires a Newton--Raphson scheme to compute 
the solution at the next time step. A case which is particularly 
demanding is one in which the RHS involves an {\it integral} over a
non-linear function of the solution, involving higher derivatives. 
This situation is encountered frequently in free-surface problems 
involving surface tension \cite{HLS94}. 
In this case, the Newton--Raphson scheme requires the inversion of a 
full matrix (as opposed to a band matrix in the case of local equations), 
which is very costly numerically. 

To cope with these challenges and to achieve stability, it has been 
recognized that it is sufficient to only treat the highest order
derivatives implicitly, the remainder can be treated explicitely. 
For example, one splits up the RHS into the sum of a lower-order 
nonlinear operator, and a linear operator containing the highest 
derivative. Then one can first compute the nonlinear part explicitly, 
and then solve a linear equation to add the highest derivative. 
However, such a split may be difficult or even impossible to find: 
the highest derivative may be contained in a non-linear and/or nonlocal 
expression. 

In a seminal paper, Hou, Lowengrub, and Shelley \cite{HLS94} found an ingenious 
way to separate the stiff part from the nonlocal, nonlinear operator
accounting for surface tension in several model equations for interfacial
flow. The stiff part can be written as a linear and local operator, 
so that implicit treatment is feasible. 
The point of the present paper is to demonstrate that while isolating
the stiff part is perhaps the gold standard for assuring stability, it 
is by no means necessary. Instead, {\it any} expression can be added to 
stabilize an explicit method, which need not be related to the 
original physical equation. In order not to change the original problem, 
the same expression is then subtracted (effectively adding zero). 
However, one (damping) part is treated implicitly, the other explicitely. 

Although this observation might seem surprising, the reason our scheme works is
explained by the very nature of implicit schemes. In an implicit scheme,
a wide range of time scales below the physical scale is not resolved,
while preserving stability. As a result, a rough model of these 
rapidly decaying modes is entirely sufficient, without loss of accuracy. 
This implies an extraordinary freedom in using higher-order derivatives 
to achieve stability, a fact that up to now does not appear to have been 
appreciated. 
The gist of this method was first proposed by Douglas and Dupont 
\cite{DD71}, to assure stability for a nonlinear diffusion equation
on a rectangle. In the present paper, we will confine ourselves to 
one spatial dimension. However, the idea easily carries over to any
spatial dimension. 

In particular, we show that stability can always be achieved, by 
adding a sufficiently large stiff contribution, which is treated
implicitly. If this additional contribution has the same short-wavelength
scaling as the stiff part of the original contribution, stabilization 
can be achieved essentially without introducing any additional error. 
However, even if the stabilizing part has a very different scaling, 
(for example a fourth-order operator being stabilized using a 
2nd-order operator), we show that the effectiveness of schemes can be much 
improved over explicit methods. 

We illustrate these points with a number of explicit examples, with
2nd, 3rd, and 4th order as the highest spatial derivative. 
First, we consider axisymmetric surface diffusion, whose RHS contains a 
nonlinear operator which is of 2nd order in the spatial derivatives. 
This equation is stabilized using a linear diffusion operator with
constant coefficients. Next, we consider one of the non-linear, non-local 
problems treated in \cite{HLS94}: the flow in a Hele-Shaw cell with surface 
tension. We treat it using a negative definite third-order operator.
Finally, we consider the Kuramoto--Sivashinsky equation, 
which is a well-known prototype of a stiff equation, since the 
highest derivative is of fourth order. To highlight the flexibility of 
our method, we show that the fourth order operator can in fact be
stabilized using ordinary diffusion, albeit at the cost of a somewhat 
increased error. 


\section{The main idea}
\label{sec:idea}
Let us illustrate our method with a non-linear diffusion 
equation in one dimension:
\beq
u_t = (D(u) u_x)_x = D'(u) u_x^2 + D(u) u_{xx},
\label{diff1}
\eeq
where the diffusion coefficient is some function of $u$. 
To treat (\ref{diff1}) implicitly, one has to solve a non-linear 
system of equations for the solution at the new time step. However, 
realizing that instability arises from the short-wavelength 
contributions, and representing $u$ as Fourier modes $w\equiv u_k$, 
effectively we have to deal with the ordinary differential 
equation 
\beq
\frac{dw}{dt} = -D(u) k^2 w \equiv -a w.
\label{ode}
\eeq
where $a > 0$. 
\subsection{Stability analysis of one Euler step}
If one solves (\ref{ode}) using an explicit (forward) 
Euler step, then between $t^n$ and $t^{n+1}=t^n+\delta t$
one arrives at
\beq
w_{n+1} = w_n - a w_n\delta t, 
\label{Euler}
\eeq
where $a > 0$. 
This iteration is unstable (diverges to infinity) if 
$|1-a\delta t| > 1$, which means that $\delta t$ must 
satisfy $\delta t < 2/a$. Remembering that the largest 
wavenumber $k$ scales like $k \approx \delta x^{-1}$, 
one arrives at the stability requirement 
\[
\delta t \lesssim \frac{2 \delta x^2}{D},
\]
which is the scaling (\ref{stab_explicit}) for the diffusion 
equation ($m=2$) (a more precise analysis, based on the 
spatial discretization of $u_{xx}$, yields 
$\delta t < \delta x^2/2D$ as the stability constraint \cite{NR07}). 

To avoid the constraint on $\delta t$, in an implicit (backward) 
Euler step the RHS of (\ref{ode}) is evaluated at $t^{n+1}$,
leading to an iteration which is {\it unconditionally stable}. 
Instead, we want to stabilize (\ref{Euler}) by adding a new piece 
$-bt$ to the RHS of (\ref{ode}), and subtracting it again. If
the first part is treated implicitly, and the second explicitely, 
(\ref{ode}) becomes:
\[
\frac{w_{n+1} - w_n}{\delta t} = -aw_n - bw_{n+1} + bw_n,
\]

which is the iteration
\beq
w_{n+1} = \left( 1 - \frac{a \delta t}{1+b \delta t} \right) w_n 
\equiv \xi(\delta t) w_n.
\label{It}
\eeq
Thus the condition for stability is $|\xi(\delta t)| < 1$ or
\beq
\left|1+(b-a)\delta t\right| < \left|1+b\delta t\right|. 
\label{stabODE}
\eeq

Let us assume for the moment the more general case where 
$a = (a_r + ia_i) \in \mathbb{C}$, with $a_r\ge0$. Then
writing $b=\lambda a$, with $\lambda \in \mathbb{R}$, (\ref{stabODE}) 
is satisfied if~:
\beq
(2\lambda-1)\left| a \right|^2 \delta t^2 + 2 a_r \delta t > 0, 
\label{stabODE1}
\eeq
which is always true if
\beq
\lambda > \frac12. 
\label{stabODE2}
\eeq
Therefore, if condition (\ref{stabODE2}) is satisfied (which is 
$b > a/2$ for $a$ real), the scheme (\ref{It}) is unconditionally stable.

In analogy to this result, in \cite{DD71} it was shown that the 
nonlinear diffusion equation (\ref{diff1}) could be stabilized 
by adding and subtracting a diffusion equation with {\it constant}
coefficient $\overline{D}$. The scheme was shown to be unconditionally
stable if 
\[
D(u) \le \frac{\overline{D}}{2}, 
\]
which corresponds exactly to the condition (\ref{stabODE2}). 

The error of an ordinary Euler step is $E \approx a \delta t^2$, so the 
cumulative error after integrating over a finite interval is proportional 
to $\delta t$, making the scheme convergent. On the other hand, since
$w_{n+1} = w_n + O(\delta t)$, the extra contribution introduced by 
the stabilizing correction is proportional to $b \delta t^2$. This 
means that as long as $b$ is of the same order as $a$, the error 
introduced is not larger than that incurred by the Euler step itself. 

\subsection{Richardson extrapolation}

In order to achieve second-order accuracy in time, we compute two 
different approximate solutions~: the first one is $w_{n+1}^{(1)}$ for 
one step $\delta t$, the second is $w_{n+1}^{(2)}$ for two half steps 
of size $\delta t/2$. Extrapolating towards $\delta t=0$ \cite{NR07},
we find the following approximate solution~:
$$
w_{n+1} = 2w_{n+1}^{(2)} - w_{n+1}^{(1)} + \mathcal{O}(\delta t^3),
$$
since the $\mathcal{O}(\delta t^2)$ error terms cancel. 
The method is therefore of second order, in the sense that the accumulated 
error scales like $\mathcal{O}(\delta t^2)$.

The stability criterion (\ref{stabODE2}) is slightly modified when using 
Richardson extrapolation. 
Using equation (\ref{It}), the full Richardson step reads~:
\beq
w_{n+1} = \left[ 2 \xi^2\left(\frac{\delta t}{2}\right)-\xi(\delta t) 
\right] w_n = \xi_R w_n.
\label{Richs}
\eeq
For simplicity, assume that both $a>0$ and $b>0$ are real. 
Then $|\xi_R|<1$ is equivalent to 
\beq
b > \frac{a\delta t - 4 + \sqrt{(a\delta t-2)^2+2a \delta t}}{3\delta t},
\label{stabR}
\eeq
which is satisfied for any $\delta t$ if
\beq
b > \frac{2a}{3},
\label{stabR1}
\eeq
making the scheme unconditionally stable. Note that in the limit
$\delta t\rightarrow \infty$, the amplification factor becomes
\[
\xi_R = 1 - \frac{3a}{b^2}\left(b-\frac{2a}{3}\right), 
\]
which remains damping as long as (\ref{stabR1}) is satisfied. 
This is an advantage over the popular Crank-Nicolson scheme, 
for which the amplification factor approaches unity if too 
large a time step is taken, leading to undamped numerical 
oscillations \cite{Richtmyer-Morton67}.

To analyze the accuracy of the scheme, we compare the result of 
(\ref{Richs}) to the exact solution, which is 
\[
w_{n+1} = w_n e^{-a\delta t}.
\]
Defining the error $E$ as 
\beq
w_{n+1} = w_n e^{-a\delta t} + w_n E,
\label{error_def}
\eeq
we find 
\beq
E = \xi_R - e^{-a\delta t} = 
\frac{\delta t^3}{6}\left(a^3 - 3ba^2 + 3ab^2  \right) 
+ O(\delta t^4).
\label{error_res}
\eeq
Thus once more if $b$ is of the same order as $a$, the error 
introduced by the extra stabilizing terms is not increased over 
a conventional second order scheme. 

To summarize, our method permits to render any explicit method 
unconditionally stable, and is second order accurate in time. 


\section{General method}
\label{sec:general}
In this section we explain the general method to be used for 
Partial Differential Equations (PDEs). The great advantage of our 
method compared for instance to the one described in \cite{HLS94} 
is that the choice of the stabilizing terms requires only a rough 
knowledge of the stiff terms in the PDE. 
We consider a partial differential equation of the form~:
\beq
\frac{\partial u}{\partial t} = f(u,t),
\label{edp}
\eeq
where $u$ is a function of space and time : $u(x,t)$. The stiffness of 
such an equation comes from the high-order space derivatives in $f(u,t)$. 
Let us consider the following discrete approximation to equation (\ref{edp}), 
between time steps $t$ and $t+\delta t$~:
\beq
\frac{u_{n+1}-u_n}{\delta t} = f(u_n,t^n) 
- \lambda \mathcal{D}[u_n] 
+ \lambda \mathcal{D}[u_{n+1}],
\label{discrete}
\eeq
where $n$ denotes the time variable ($t^n=n \delta t$) and 
$\mathcal{D}$ is a linear damping operator, with negative eigenvalues.
We will discuss other choices below, but a particular example is the 
diffusion operator
\beq
\mathcal{D}[u] = \frac{\partial^2 u}{\partial x^2} . 
\label{damping}
\eeq

In terms of the increments $\delta u=u_{n+1}-u_n$, equation 
(\ref{discrete}) reads~:
\beq
\frac{\delta u}{\delta t} = f(u_n,t^n) 
+ \lambda \mathcal{D}[\delta u].
\label{incr}
\eeq
This linear system of equations can be written in terms of a linear 
operator $\mathcal{L} = I - \lambda \delta t \mathcal{D}$~:
\beq
\mathcal{L} \cdot \delta u = f(u_n,t^n) \; \delta t,
\label{lin_eq}
\eeq
and once it has been inverted, we obtain the solution $u$ at time $t^{n+1}$~:
\beq
u_{n+1} = u_n + \mathcal{L}^{-1} \cdot f(u_n,t^n) \; \delta t.
\label{discrete_sol}
\eeq

The difficult part in this last step comes from the fact that solving 
(\ref{lin_eq}) may be computationally expensive. However, 
we will choose $\mathcal{D}$, and therefore the linear operator $\mathcal{L}$,
such that this numerical procedure only requires $\mathcal{O}(N)$, 
or at most $\mathcal{O}(N \log N)$ operations, where $N$ is the number 
of grid points.

As explained in section~\ref{sec:idea}, in order to achieve second-order 
accuracy in time, we compute two approximate solutions~: 
the first one $u_{n+1}^{(1)}$ for one step $\delta t$, 
the second one $u_{n+1}^{(2)}$ for two half steps $\delta t/2$. By 
extrapolating towards $\delta t=0$, we find the following approximate 
solution~:
$$
u_{n+1} = 2u_{n+1}^{(2)} - u_{n+1}^{(1)} + \mathcal{O}(\delta t^3),
$$
where the $\mathcal{O}(\delta t^2)$ error terms cancel. The method is 
therefore second order, in the sense that cumulated errors on a number 
of time steps of the order of $1/\delta t$ are $\mathcal{O}(\delta t^2)$.

\subsection{Choice of damping operator}
\label{sub:damping}

The crucial step of our method consists in choosing the right 
damping operator $\mathcal{D}$. Since its only purpose is to damp the 
high-order derivative on the RHS of the PDE, it does not need to be 
computed with great accuracy, but only needs to have the same scaling 
in wavenumber than the stiff term in the PDE, as we will explain below. 

We derive the scaling for the critical value of $\lambda$ and for
the error, both of which will be calculated more precisely for each 
of the individual examples below. We look for solutions with 
a single Fourier mode of the form $h_j^{n} = \xi^n e^{ik x}$, 
where $\xi(\delta t,k)$ is the amplification factor \cite{NR07}. 
We assume that 
\beq
\mathcal{D}[e^{ikx}] = -|k|^d e^{ikx}, 
\label{D_scal}
\eeq
and so $d = 2$ in the case of the diffusion operator. Inserting this 
into equation (\ref{incr}), we obtain after simplification~:
$$
\frac{\xi-1}{\delta t} = \frac{f(\xi^n e^{ik x},t^n)}{\xi^n}
-\lambda |k|^d (\xi-1).
$$

Suppose now that the function $f$ contains a stiff linear term, 
{\em i.e.} with several space derivatives~:
$$
f(\xi^n e^{ik x},t^n) \sim - A \xi^n |k|^m,
$$
where $A$ is a positive constant.
Then the amplification factor reads~:
\beq
\xi(\delta t,k) = 1 - \frac{A |k|^m \delta t}{1 + \lambda |k|^d \delta t},
\eeq
and comparing to (\ref{It}) we can identify:
\beq
a = A|k|^m, \quad b = \lambda |k|^d.
\eeq
To investigate stability, we must consider the largest possible
wave number, which is of order $\delta x^{-1}$. This means for both the 
Euler and the Richardson scheme, stability is guaranteed for 
\beq
\lambda \gtrsim A \delta x^{d-m}. 
\label{lambda_est}
\eeq

Thus our first observation is that we can {\it always} stabilize 
an explicit method with our scheme, even if the stabilizing operator
is of lower order. We will demonstrate this explicitely when we 
discuss the Kuramoto--Sivashinsky equation below. However, it is 
{\it preferable} to choose $d = m$ (so that $\lambda$ becomes independent 
of $\delta x$), since otherwise the error increases over that of 
a fully implicit method. Namely, as we have seen in the previous section, 
the error introduced by the stabilizing term is $b\delta t$ or 
$(b\delta t)^2$ for the Euler and Richardson schemes, respectively.

Now let $\Delta$ be the size of the physical scale that needs to 
be resolved, which means we have to guarantee $b\delta t \lesssim 1$ 
for $k \approx \Delta^{-1}$. Thus using $\lambda$ as estimated by 
(\ref{lambda_est}), we find that the time step has to satisfy the 
constraint 
\beq
\delta t \lesssim \frac{\delta x^m}{A}\left(\frac{\Delta}{\delta x}\right)^d,
\label{constraint}
\eeq
to achieve reasonable accuracy on scale $\Delta$. 
Now comparing to (\ref{stab_explicit}) there is an improvement given 
by the factor $(\Delta/\delta x)^d$ on the right. If $d = m$, the scaling
is the same as for an implicit method, for which the error is determined 
by the physical scale $\Delta$ alone. 

To satisfy the condition $d = m$, various types of stabilizing 
operators $\mathcal{D}[u]$ need to be considered. In choosing 
$\mathcal{D}[u]$, we also need to ensure that equation (\ref{lin_eq}) 
can be inverted with the fewest number of numerical operations. 
The following cases can be encountered~:
\begin{itemize}
\item The simplest case is obtained when $m=2$, in which case the 
diffusion operator (\ref{damping}) ensures $d = m$. 
This operator, when discretized using centered finite differences,
leads to a tridiagonal matrix and equation (\ref{lin_eq}) can be 
solved in $\mathcal{O}(N)$ operations. 
\item If $m=4$, a fourth-order diffusion operator 
$\mathcal{D}[u]=-\partial^4 u / \partial x^4$ ensures $d = m$, and 
leads to a penta-diagonal system that can be solved in $\mathcal{O}(N)$ 
operations. 
\item When $m$ is odd, for instance $m=3$, as it will be shown to be the 
case for Hele-Shaw flows, an $m^{th}$-order space derivative 
does not correspond to a damping operator, but to a traveling wave 
term (purely imaginary term in Fourier space). 
In order to achieve the right scaling for $|k|\gg1$, we use 
$\mathcal{D}[u]=\mathcal{H}[\partial^3 u / \partial x^3]$, 
where $\mathcal{H}$ is the Hilbert transform. This expression has the 
correct scaling $\mathcal{H}[\partial^3 u / \partial x^3] \sim -k^3$, 
and can be inverted very easily in Fourier space in 
$\mathcal{O}(N \log N)$ operations.
\end{itemize}

In conclusion, any high-order space derivative in the RHS of the 
PDE can be stabilized using an {\em ad hoc} operator $\mathcal{D}[u]$, 
using a number of numerical operations at most equal to $\mathcal{O}(N \log N)$.


\section{Mean curvature flow}

\subsection{Numerical scheme}

Axisymmetric motion by mean curvature \cite{EF09} is described by the 
equation~:
\beq
h_t = \frac{h_{xx}}{1+h_x^2} - \frac{1}{h}, 
\label{mc}
\eeq
where $h(x,t)$ is the local radius of a body of revolution. Geometrically,
it describes an interface motion where the normal velocity of the 
interface is proportional to the mean curvature. Physically, 
(\ref{mc}) describes the melting and freezing of a $^3He$ crystal, 
driven by surface tension \cite{IGRBE07}. Generic initial conditions
lead to pinch-off in finite time \cite{DK91}, and thus require high 
demands on the resolution and stability of the numerical method. 
In our example, we will use Dirichlet boundary conditions with a 
periodic initial perturbation:
$$
h(0,t)=h(L,t)=1, \qquad h(x,0)=1+0.1 \sin(2\pi x /L),
$$
where $L=10$.
As seen in Fig.~\ref{profiles_mc}, this initial condition leads
to pinch-off in finite time. 

Owing to the second derivative on the RHS, (\ref{mc}) is numerically
stiff. However, since there is a nonlinear term multiplying $h_{xx}$, 
an implicit scheme requires the solution of a nonlinear equation \cite{BBW98}. 
We will stabilize the stiff part of the equation by adding and subtracting
a term $\lambda h_{xx}$, where $\lambda$ needs to be determined. 
The resulting equation is discretized on a regular grid, using centered 
finite differences~:
\beq
\frac{h_j^{n+1}-h_j^n}{\delta t} = \frac{2}{\delta x} \cdot 
\frac{h_{j-1}^n-2h_j^n+h_{j+1}^n}{2\delta x + h_{j+1}^n-h_{j-1}^n}
- \frac{1}{h_j^n}
- \lambda \frac{h_{j-1}^n-2h_{j}^n+h_{j+1}^n}{\delta x^2}
+ \lambda \frac{h_{j-1}^{n+1}-2h_{j}^{n+1}+h_{j+1}^{n+1}}{\delta x^2}. 
\label{scheme_mc}
\eeq

\subsection{Von Neumann stability analysis}

Using a ``frozen coefficient'' hypothesis, we look for perturbations 
to the mean profile $\overline{h}$ in the form of a single Fourier mode:
\[
h_j^{n} = \overline{h}(j\delta x,n\delta t)+ 
\xi^n e^{ik j \delta x},
\]
where $\xi(\delta t,k)$ is the amplification factor \cite{NR07}. 
Inserting this expression into equation (\ref{scheme_mc}) and 
linearizing in the perturbation, we obtain after some simplification~:
$$
\frac{\xi-1}{\delta t} = \frac{2}{\delta x^2\left(1+\overline{h}_x^2\right)} 
\left( \cos(k \delta x) -1 \right) - \frac{1}{\overline{h}^2} + 
\frac{2 \lambda}{\delta x^2} (\xi -1) \left( \cos(k \delta x) -1 \right).
$$
Using that $\delta x \ll \overline{h}$, 
we can identify coefficients $a$ and $b$ from equation (\ref{It})~:
\beq 
a = \frac{2}{\delta x^2\left(1+\overline{h}_x^2\right)} 
\left( 1 - \cos(k \delta x) \right) \quad \mathrm{and} \quad 
b = \frac{2 \lambda}{\delta x^2} \left( 1 - \cos(k \delta x) \right).
\eeq
We have shown that the Richardson extrapolation scheme is stable if 
$b>2a/3$, which implies~:
\beq
\lambda > \frac{2}{3\left(1 + \overline{h}_x^2\right)}.
\label{stab_mc}
\eeq
For the simulation to be reported below, we take $\lambda=0.7$, which 
satisfies (\ref{stab_mc}) uniformly, and makes our numerical scheme 
unconditionally stable. On the other hand, we confirmed that instability
occurs if we choose $\lambda \le 0.5$, so that (\ref{stab_mc}) is
violated over some parts of the solution. 

\begin{figure}[h!]
\begin{center}
\includegraphics[width=0.8\linewidth]{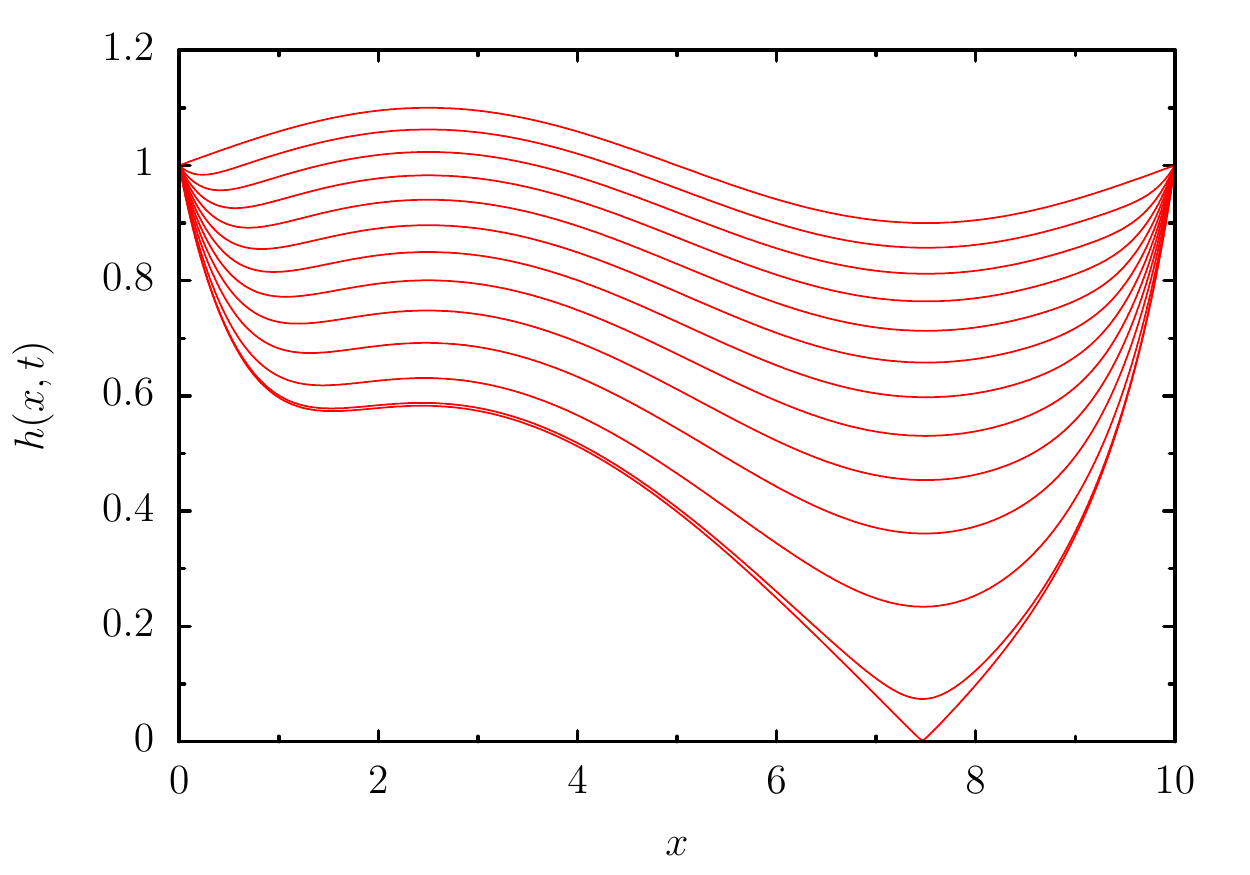}
\end{center}
\caption{Successive profiles of the interface described by
(\ref{mc}), with $\lambda = 0.7 > 2/3$. The criterion (\ref{stab_mc})
for unconditional stability is verified uniformly. 
}
\label{profiles_mc}
\end{figure}

The linear tridiagonal system (\ref{scheme_mc}) is solved for $h^{n+1}$ 
and Richardson extrapolation is used to obtain second-order 
accuracy in time. Figure~\ref{profiles_mc} shows successive profiles 
of the interface, until a minimum height of $h_{min} = 10^{-3}$ is reached.
We have used a uniform grid with $N=2048$ grid points. Since the
characteristic time scale of the solution goes to zero as pinch-off is 
approached, we use a simple adaptive scheme for the time step~:
a relative error is computed using the two estimates 
of the solution for $\delta t$ and $\delta t/2$; then if this error is 
larger than $10^{-5}$, the time step is divided by $2$.

\begin{figure}[h!]
\begin{center}
\includegraphics[width=0.8\linewidth]{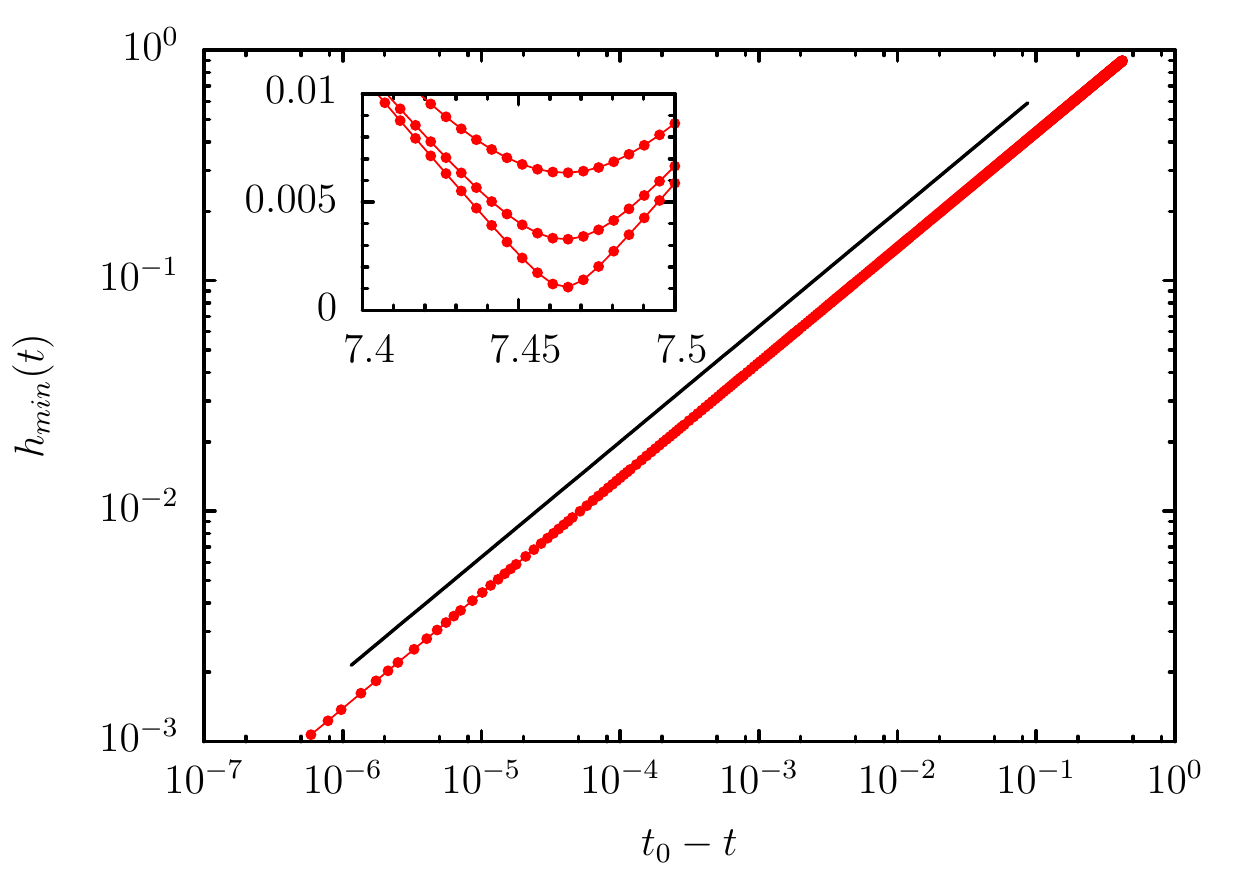}
\end{center}
\caption{Minimum radius of the interface as a function of $t_0-t$, where 
$t_0$ is the pinch-off time. The full line corresponds to $(t_0-t)^{1/2}$.
In the inset, we show a closeup of near the pinch point, plotting 
a profile every 10 time steps (but with adaptive time step). 
}
\label{rmin}
\end{figure}

Close to pinch-off, (\ref{mc}) exhibits type-II self-similarity 
\cite{EF09}, characterized by the presence of logarithmic corrections
to power law scaling. However, the minimum radius $h_{min}$ scales 
with a simple power law exponent of $1/2$. To test this, and to confirm
stability of our scheme down to very small scales, we followed the 
solution until spatial resolution was lost. In Fig.~\ref{rmin},
we show a doubly logarithmic plot of $h_{min}$ as a function of $t_0-t$, 
where $t_0$ is the singularity time. We determined $t_0$ by extrapolating 
$h_{min}(t)$ towards zero. The numerical solution is seen to exhibit the 
expected scaling down to the smallest resolvable scales, as illustrated
in the inset. 

To confirm that the method is indeed second-order accurate in time, 
we calculated the error as the $\infty-$norm of the difference between a 
numerical solution at a fixed time $t=0.4$, 
obtained with time step $\delta t=0.4\times 2^{-m}$ 
and the "exact" solution obtained with the smallest time step 
$\delta t_{min}=0.4\times 2^{-16}$, divided by the maximum value of 
the solution~:
\beq
\textrm{Error}=\frac{Max_j \left| f_j(\delta t_{min}) - 
f_j(\delta t)\right|}{Max_j \left| f_j(\delta t_{min})\right|}.
\label{Error_MC}
\eeq
\begin{figure}[h!]
\begin{center}
\includegraphics[width=0.8\linewidth]{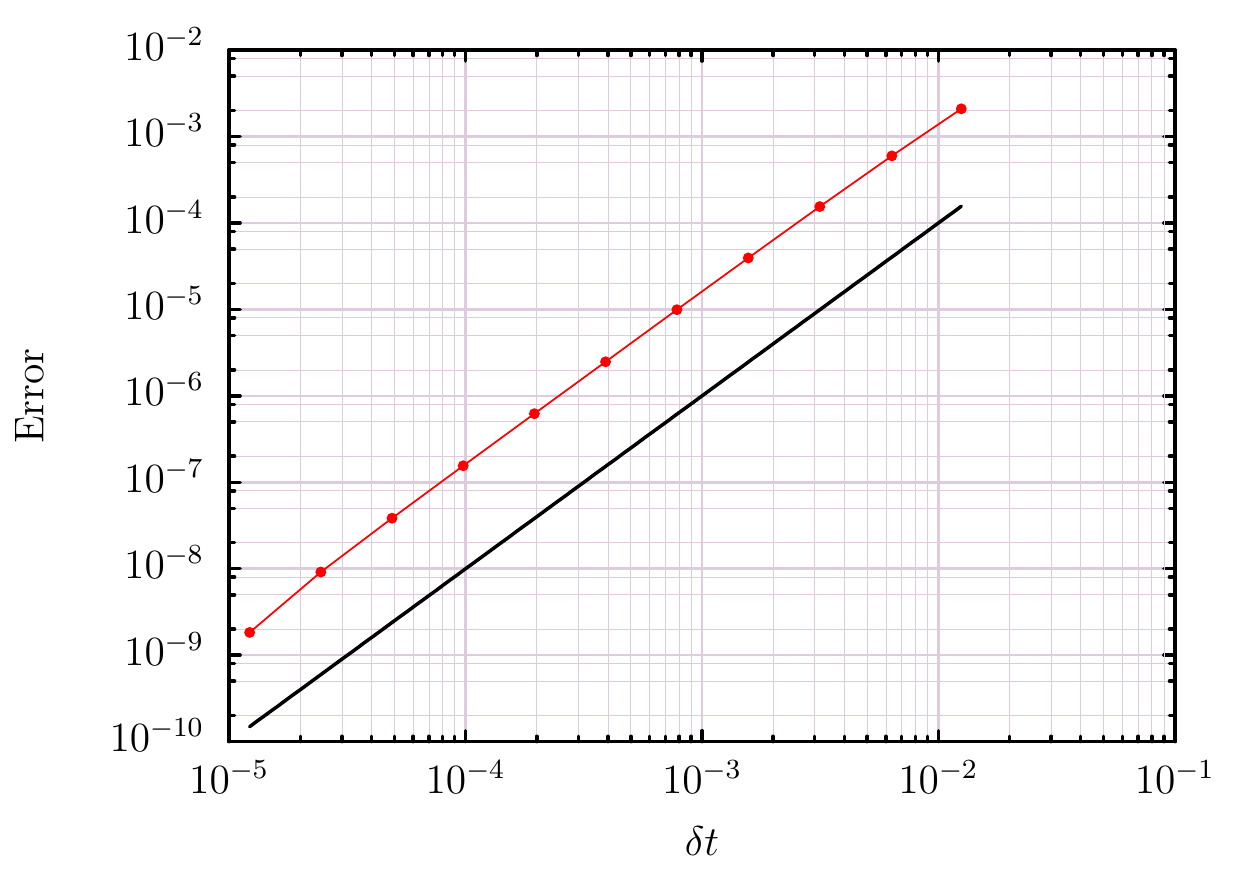}
\end{center}
\caption{The maximum error (\ref{Error_MC}) of the numerical solution 
of (\ref{mc}) at $t=0.4$, as a function of $\delta t$. The solid line 
corresponds to a quadratic dependence, which shows that the error 
using Richardson extrapolation scales like $\delta t^2$.}
\label{error_dt_mc}
\end{figure}
As seen in Fig.~\ref{error_dt_mc}, the error scales indeed like
$\delta t^2$, the same as expected for a fully implicit method, 
using for example the Crank-Nicolson scheme.


\section{Hele-Shaw flow with surface tension}

In the preceeding example, a fully implicit treatment of the right
hand side of (\ref{mc}) would at least have been feasible \cite{BBW98},
although our method simplifies the algorithm. By contrast, examples
presented in \cite{HLS94} are much more challenging, in that the RHS
is both nonlinear and non-local. We focus on the particular example 
of Hele-Shaw interface flow, whose spatial derivatives scale like 
$|k|^3$ in Fourier space, and thus lead to a very stiff system.
While it is perfectly possible to stabilize 
the scheme using ordinary diffusion, according to our analysis of 
subsection \ref{sub:damping}, this would entail an increased time
truncation error. Therefore, we use the Hilbert transform to construct
a stabilizing term which is of third order in the spatial derivative, 
yet only takes $\mathcal{O}(N \log N)$ operations to calculate. 

\subsection{Equations}

We consider an interface in a vertical Hele-Shaw cell, separating 
two viscous fluids with the same dynamic viscosity \cite{HLS94}.
As heavy fluid falls, a slightly perturbed interface is deformed strongly 
by gravity, while surface tension assures regularity on small scales, 
as seen in Fig.~\ref{fig:evol1D} below. For simplicity, we assume 
the flow to be periodic in the horizontal direction. 

The interface is discretized using marker points labeled with $\alpha$, 
which are advected according to~:
\beq
\frac{\partial{\bf X}(\alpha)}{\partial t} = U {\bf n} + T {\bf s}.
\label{advection}
\eeq
Here ${\bf X}(\alpha)=(x,y)$ is the position vector, 
${\bf n}=(-y_\alpha/s_\alpha,x_\alpha/s_\alpha)$ and 
${\bf s}=(x_\alpha/s_\alpha,y_\alpha/s_\alpha)$ 
are the normal and tangential unit vectors, respectively. 
Hence $U=(u,v)\cdot {\bf n}$ and $T=(u,v)\cdot {\bf s}$ 
are the normal and tangential velocities, respectively. 
Since the evolution of a surface is determined only by its normal 
velocity, we can choose the tangential velocity of the marker points freely, 
in order to keep a reasonable distribution of points and avoid point clustering.
The precise choice of the tangential velocity will be described later.
For an unbounded interface, the complex velocity of marker points 
labeled with $\alpha$ is given by the Birkhoff--Rott 
integral \cite{MB02}:
\beq
w(\alpha)=u(\alpha)-i v(\alpha)=\frac{1}{2 \pi i} 
PV\int_{-\infty}^{+\infty}{\frac{\gamma(\alpha',t)}
{z(\alpha,t)-z(\alpha',t)} d\alpha'},
\label{w_int}
\eeq
where $z(\alpha,t)=x+iy$. Here $\gamma$ is the vortex sheet strength at
the interface. If the surface is periodic with period $1$ 
($z(\alpha+2\pi)=z(\alpha)+1$), (\ref{w_int}) can be written as an 
integral over the periodic domain $\alpha\in [0,2\pi]$ of the label:
\beq
u(\alpha)-i v(\alpha)=\frac{1}{2i} 
PV\int_{0}^{2\pi}{\gamma(\alpha',t)\cot
\left[ \pi(z(\alpha,t)-z(\alpha',t))\right] d\alpha'},
\label{BR}
\eeq
where we have used the continued fraction representation of the cotangent
\cite{Carrier66}~:
$$
\pi \cot (\pi z) = \frac{1}{z} + 2 z \sum_{k=1}^\infty{\frac{1}{z^2-k^2}}.
$$

For two fluids of equal viscosity, the vortex sheet strength 
$\gamma$ is given by \cite{MB02}~:
\beq
\gamma = S \kappa_{\alpha} - R y_\alpha,
\label{gamma}
\eeq
where $\kappa$ is the mean curvature of the interface~:
\beq
\kappa(\alpha)=\frac{x_\alpha y_{\alpha \alpha}-y_\alpha x_{\alpha \alpha}}
{s_\alpha^3}, \qquad \mathrm{with} \qquad
s_\alpha=(x_\alpha^2+y_\alpha^2)^{1/2}.
\label{kappa}
\eeq
Here $S$ is the non-dimensional surface tension coefficient and $R$ is the 
non-dimensional gravity force. As an initial condition, we choose the 
same as the one used in \cite{HLS94}, which corresponds to a slight modulation 
of a flat interface:
\beq
x(\alpha,0)= \alpha/2\pi, \quad y(\alpha,0)=0.01 (\cos(\alpha)-\sin(3\alpha)).
\label{HS_init}
\eeq

To compute the complex Lagrangian velocity of the interface (\ref{BR}), 
we use the spectrally accurate alternate point discretization \cite{S92}~:
\beq
u_j - i v_j \simeq -\frac{2\pi i}{N} \sum_{{l=0}\atop{j+l \; odd}}^{N-1}
{\gamma_l \cot\left[\pi(z_j-z_l)\right]}.
\label{comp_vel}
\eeq
$\kappa_\alpha$ and $y_\alpha$ are computed at each time step using 
second-order centered finite differences, and $\alpha$ is defined by 
$\alpha(j) = 2 \pi j/N$, where $j \in [0,N]$ and $N$ is the number of 
points describing the periodic surface. Note that the numerical effort
of evaluating (\ref{comp_vel}) requires $\mathcal{O}(N^2)$ operations,
and thus will be the limiting factor of our algorithm. 
For the tangential velocity $T$, 
we use the same expression as \cite{HLS94}, which is designed to avoid 
point clustering. For completeness, we describe the procedure in the
Appendix. 

\subsection{Third-order stabilizing operator}

It follows from (\ref{BR}) and (\ref{gamma}), 
that the Hele-Shaw dynamics contains a stiff part which scales like 
$|k|^3$ in Fourier space \cite{HLS94}. As a result, we need to define a 
third-order 
operator to stabilize the equations. When the interface is described 
using marker points labeled with $\alpha$, the most natural choice of 
damping operating on the Cartesian coordinates 
$\left(x(\alpha),y(\alpha)\right)$ is~:
\beq
\mathcal{D}\left[\left(x(\alpha),y(\alpha)\right)\right] = 
\left(\mathscr{H}(x_{\alpha \alpha \alpha}),
\mathscr{H}(y_{\alpha \alpha \alpha})\right).
\label{op_HS}
\eeq
Here $\mathscr{H}$ is the Hilbert transform~:
$$
\mathscr{H}\left[ f \right](\alpha) = 
\frac{1}{\pi} \int_{-\infty}^{+\infty}{\frac{f(\alpha')}
{\alpha-\alpha'}d\alpha'},
$$
which satisfies~:
\beq
\mathscr{H}\left[ e^{i k x} \right] = -i \; 
\mathrm{sign} (k) \; e^{i k x}, \qquad \mathscr{H}\left[ 1 \right] = 0. 
\label{hilb_prop}
\eeq
Note that 
\[
R \mathcal{D}\left[\left(x(\alpha),y(\alpha)\right)\right] = 
\mathcal{D} R \left[\left(x(\alpha),y(\alpha)\right)\right],
\]
where $R$ is an arbitrary rotation matrix. Thus the stabilizing
terms share the same invariance under rotation as the original 
problem (\ref{advection}). 

Using the first property (\ref{hilb_prop}), one notes that the scaling 
of the operators for a single mode $e^{i k \alpha}$ 
(in $x$ or $y$) is~:
\beq
\mathcal{D} [e^{i k \alpha}] = - \left| k \right|^3 e^{i k \alpha}.
\label{D_op}
\eeq
Using the representation (\ref{D_op}) in Fourier space, 
we can compute the stabilizing operators (\ref{op_HS}) in 
$\mathcal{O}(N \log N)$ operations with the aid of the
Fast Fourier Transform \cite{FFTW05}, which
leads to the following numerical algorithm.

First, 
the set of horizontal coordinates of the marker points has to be
modified, such that $x'_j$ is periodic:
$$
x'_j = x'(\alpha(j)) = x(\alpha(j)) - j /N.
$$
Now we are able to compute the discrete Fourier transform 
of $x'_j$:
\beqa
\hat{x}'_k & = & \sum_{j=0}^{N-1}{x'_j e^{-2 i \pi j k / N}},
\eeqa
where $k = 0, \dots, N-1$. Modes with $k>N/2$ correspond to negative
wavenumbers of modulus $N-k$. 
The transforms of $y$, as well as the velocity components $u$ and 
$v$ are defined analogously. Each of these transforms can be performed 
using the Fast Fourier Transform, the velocity components at the 
old time step $n$ are computed from (\ref{comp_vel}). 

Now the discrete version of (\ref{advection}) becomes, including the 
stabilizing terms~:
\beqa
\frac{\hat{x}'^{n+1}_k-\hat{x}'^{n}_k}{\delta t} & = & 
\hat{u}_k^n - \lambda k^3 \hat{x}'^{n+1}_k + \lambda k^3 \hat{x}'^{n}_k
\label{y_equ1}\\
\frac{\hat{y}_k^{n+1}-\hat{y}_k^{n}}{\delta t} & = & 
\hat{v}_k^n - \lambda k^3 \hat{y}_k^{n+1} + \lambda k^3 \hat{y}_k^{n},
\label{y_equ2}
\eeqa
where $\hat{x}'_k$ and $\hat{y}_k$ are complex numbers, and 
$k = 0, \dots, N/2$. From equations (\ref{y_equ1}) and (\ref{y_equ2}), 
$\hat{x}'^{n+1}_k$ and $\hat{y}_k^{n+1}$ are found according to~:
\beqa
\label{it1}
\hat{x}'^{n+1}_k & = & \hat{x}'^{n}_k + \frac{\hat{u}_k^n \; \delta t}
{1+\lambda k^3 \delta t}\\
\label{it2}
\hat{y}_k^{n+1} & = & \hat{y}_k^{n} + \frac{\hat{v}_k^n \; \delta t}
{1+\lambda k^3 \delta t}.
\eeqa
For $k > N/2$, Fourier coefficients are found from 
$\hat{x}'^{n+1}_k = \left(\hat{x}'^{n+1}_{N-k}\right)^*$ and  
$\hat{y}^{n+1}_k = \left(\hat{y}^{n+1}_{N-k}\right)^*$. 
Finally, the inverse Fourier Transform of $\hat{x}'^{n+1}_k$ and 
$\hat{y}_k^{n+1}$ yields the components of $x$ and $y$ at the 
new time step:
\beqa
x_j^{n+1} & = & \frac{1}{N} \sum_{k=0}^{N-1}{\hat{x}'^{n+1}_k 
e^{2 i \pi j k / N}} + \frac{j}{N}\\
y_j^{n+1} & = & \frac{1}{N} \sum_{k=0}^{N-1}{\hat{y}_k^{n+1} 
e^{2 i \pi j k / N}}.
\eeqa
The cost of this procedure represents a small effort compared to the 
evaluation of the velocities (\ref{comp_vel}), which requires
$\mathcal{O}(N^2)$ operations.

\subsection{von Neumann stability analysis}
We are considering the amplification of small short-wavelength 
perturbations on the interface. In view of the rotational invariance 
of the system of equations, we can suppose an almost horizontal interface
$0\le x \le 1$:
\[
z_j = \frac{j}{N} + i y_j,
\]
where the $y_j$ are small and represent small perturbations. Then the 
linearization of (\ref{gamma}) and
(\ref{kappa}) reads (keeping only the highest derivative):
\[
\gamma_j^n = S \frac{\kappa^n_{j+1}-\kappa^n_{j-1}}{2\delta\alpha}, 
\quad \kappa^n_j = (2\pi)^2\frac{y^n_{j+1}-2y^n_j+y^n_{j-1}}{\delta\alpha^2},
\]
and the explicit part of the equation is, using (\ref{comp_vel}):
\beq
\frac{y_j^{n+1}-y_j^n}{\delta t} = 
\frac{2\pi}{N} \sum_{{\ell=0}\atop{j+\ell \; odd}}^{N-1}
{\gamma^n_\ell \cot\left[\frac{\pi}{N}(j-\ell)\right]}.
\label{dis_lin}
\eeq

As before, we make the ansatz $y_j^n = \xi^n e^{ikj\delta\alpha}$, 
which yields 
\[
\gamma^n_j = 2i\xi^n S\frac{(2\pi)^2}{\delta\alpha^3}
\left(\cos\frac{2\pi k}{N}-1\right)\sin\frac{2\pi k}{N}e^{ikj\delta\alpha}.
\]
Finally, using the discrete form of the Hilbert transform, we 
obtain \cite{C70}:
\beq
\frac{\xi-1}{\delta t} = 2i S\frac{(2\pi)^3}{\delta\alpha^3}
\left(\cos\frac{2\pi k}{N}-1\right)\sin\frac{2\pi k}{N}H_k,
\label{H_discr}
\eeq
where 
\[
H_k=   \left\{\begin{array}{l}
         -i/2  \quad 1\le k\le N/2-1 \\
           0     \quad k=0,N/2 \\
          i/2 \quad N/2+1 \le k \le N-1\quad\;.
                 \end{array}\right.
\]

Thus for wavenumbers $1\le k \le N/2-1$, 
we can identify the coefficient $a$ in equation (\ref{It})
as 
\beq
a = S N^3\left(\cos\frac{2\pi k}{N}-1\right)\sin\frac{2\pi k}{N}.
\eeq
On the other hand, $\hat{y}_k^n = N \xi^n$, so according to (\ref{y_equ1}) 
and (\ref{y_equ2})
we find 
\beq
b = \lambda k^3.
\eeq

The Richardson scheme will be stable if $b>2a/3$, which means that~:
\beq
\lambda > \frac{2S}{3} 
\frac{N^3}{k^3}\left(\cos\frac{2\pi k}{N}-1\right)\sin\frac{2\pi k}{N},
\label{HS_stable}
\eeq
a condition which needs to be satisfied for all $k \le N/2-1$ to 
guarantee unconditional stability. The maximum of the right hand side
of (\ref{HS_stable}) is $(2\pi)^3/2$, which is in fact achieved in the
limit of {\it small} $k$, as verified numerically below. Thus the 
stability criterion becomes 
\beq
\lambda > \frac{S(2\pi)^3}{3}\approx 82.7 S. 
\label{HS_stable_max}
\eeq

So far our calculation was based on an interface of length 1, while the
markers run from $0$ to $2\pi$; this is the origin of the factor 
$(2\pi)^3 = s_{\alpha}^{-3}$ in (\ref{HS_stable_max}). As the interface 
is stretched during the computation, $s_{\alpha}$ increases and the 
stability constraint becomes less stringent:
\beq
\lambda > \frac{S}{3 s_{\alpha}^3}. 
\label{HS_stable_gen}
\eeq
To achieve an optimal result, one could choose $\lambda$ 
as a function of space and time, depending on the local value
of $s_{\alpha}$. For simplicity, in the computations reported below,
we choose $\lambda$ to be time dependent only, based on the minimum 
value of $s_{\alpha}$ over space, which is estimated as 
\[
s_{\alpha} \approx \frac{\delta s_{min} N}{2\pi},
\]
where $\delta s_{min}$ is the minimum spatial distance between 
grid points. 

\subsection{Numerical results}
\begin{figure}[h!]
\begin{center}
\includegraphics[width=0.8\linewidth]{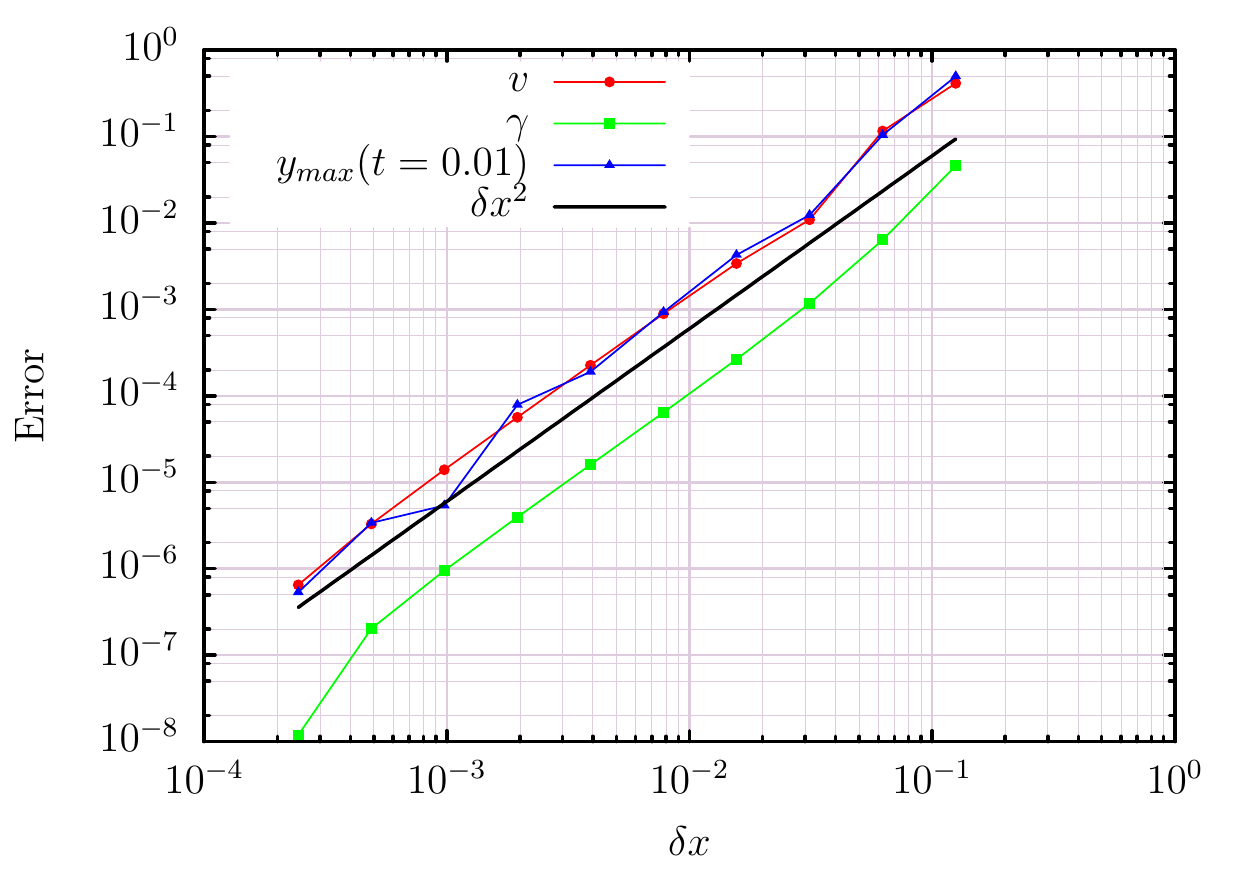}
\caption{Spatial truncation errors, as a function of 
the initial uniform $\delta x$. The ``true'' solution is obtained with the 
finest grid, with $2^{13}$ points. The red circles correspond to the 
relative error on the vertical velocity at $t=0$, computed using equation 
(\ref{BR}). The green squares correspond to the relative error on the 
vortex sheet strength $\gamma$ at $t=0$, computed using equation 
(\ref{gamma}). The blue triangles correspond to the relative error on 
the maximum value of $y$ at $t=0.01$. The solid line shows that all these 
relative errors scale like $\delta x^2$.
}
\label{error_dx_HS7}
\end{center}
\end{figure}

The spatial convergence of our method has been tested by comparing
to a "true" solution, computed using a fine grid with $N=2^{13}$. 
Then, the relative error 
to this solution is computed for the vertical velocity and the vortex 
sheet strength at $t=0$, for smaller values of $N$.
A relative error of the whole code is also computed for the height of 
the interface at $t=0.01$. Figure \ref{error_dx_HS7} presents these 
results, plotted against the initial $\delta x$, together with a power 
fit, that proves the spatial convergence to scale like $\delta x^2$. 

\begin{figure}[h!]
\begin{center}
\includegraphics[width=\linewidth]{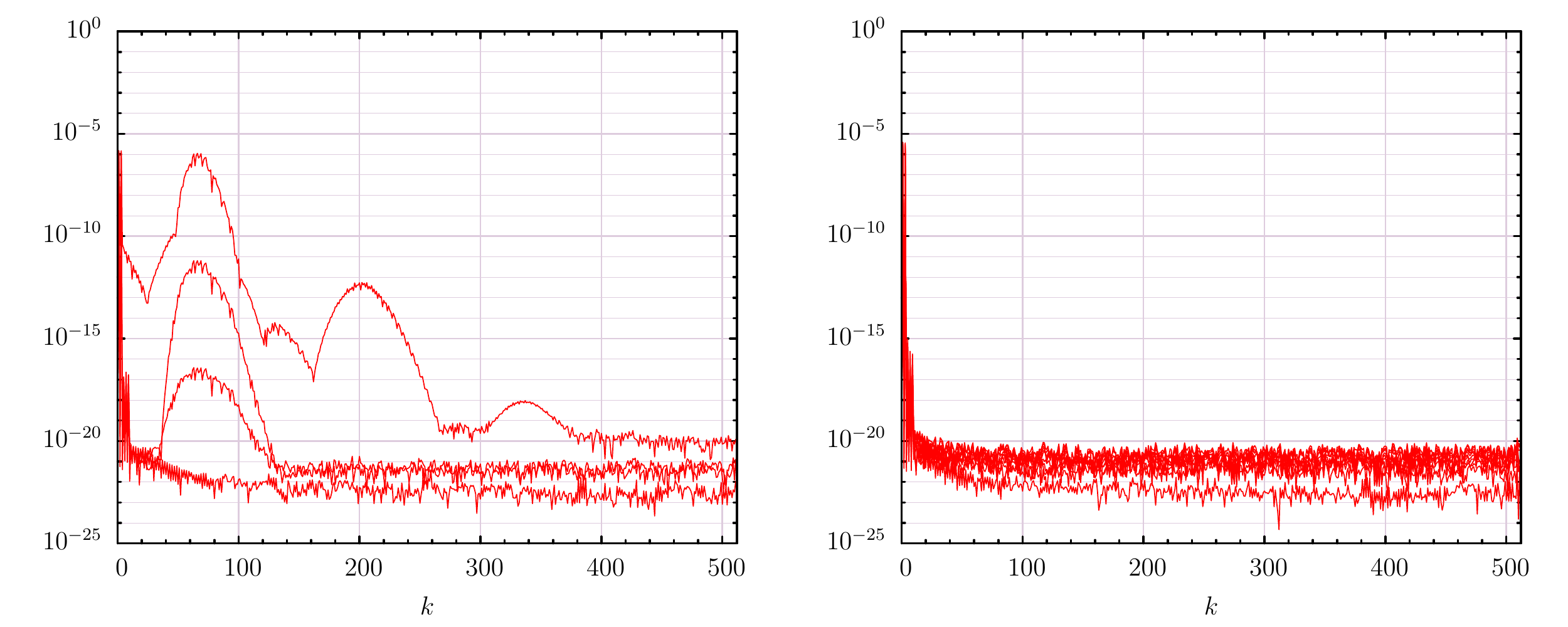}
\caption{Evolution of the amplitude spectrum of the vertical displacement 
$y_j^n$ for two different values of $\lambda$~: $\lambda=70 \; S$ on the left, 
$\lambda=85 \; S$ on the right. The initial condition for both these 
computations is the same as (\ref{HS_init}), except that the initial 
amplitude is $10^{-6}$. The left computation becomes unstable for 
$t \gtrsim 3 \times 10^{-3}$, 
whereas the right one remains stable. In both cases, $1024$ points have 
been used and the time step is $\delta t = 3.125 \times 10^{-5}$.
Spectra are shown every 30 time steps. 
}
\label{spectra}
\end{center}
\end{figure}

To probe the damping of any numerical instability by the scheme 
(\ref{it1}),(\ref{it2}), we recorded the spectra of a solution 
close to a horizontal interface, with a very small perturbation 
added to it. The spectra are shown every 30 time steps, starting 
from the initial condition. If $\lambda$ is chosen
slightly larger than the boundary (\ref{HS_stable_max}), the 
spectrum remains flat and free of unphysical growth for large wave numbers, 
as seen on the right of Fig.~\ref{spectra}. If on the other hand 
$\lambda$ is chosen somewhat smaller, numerical instability first occurs
toward the small wavenumber end of the spectrum, as seen on 
the left of Fig.~\ref{spectra}. This is in agreement with (\ref{HS_stable}),
which shows that the stability condition is first violated for small $k$. 

Although this might seem unusual at first, the observed growth
for small $k$ is simply a result of our choice of damping, which 
slightly emphasizes large wavenumbers relative to smaller ones.
Had we chosen to implement (\ref{op_HS}) using finite differencing 
for $x_{\alpha \alpha \alpha}$ and $y_{\alpha \alpha \alpha}$, 
and only performing the Hilbert transform in Fourier space, the
critical value of $\lambda$ would have been {\it independent} of 
$k$. 

\begin{figure}[hbt]
\begin{centering}
\subfigure[]
         {\includegraphics[width=0.49\linewidth]{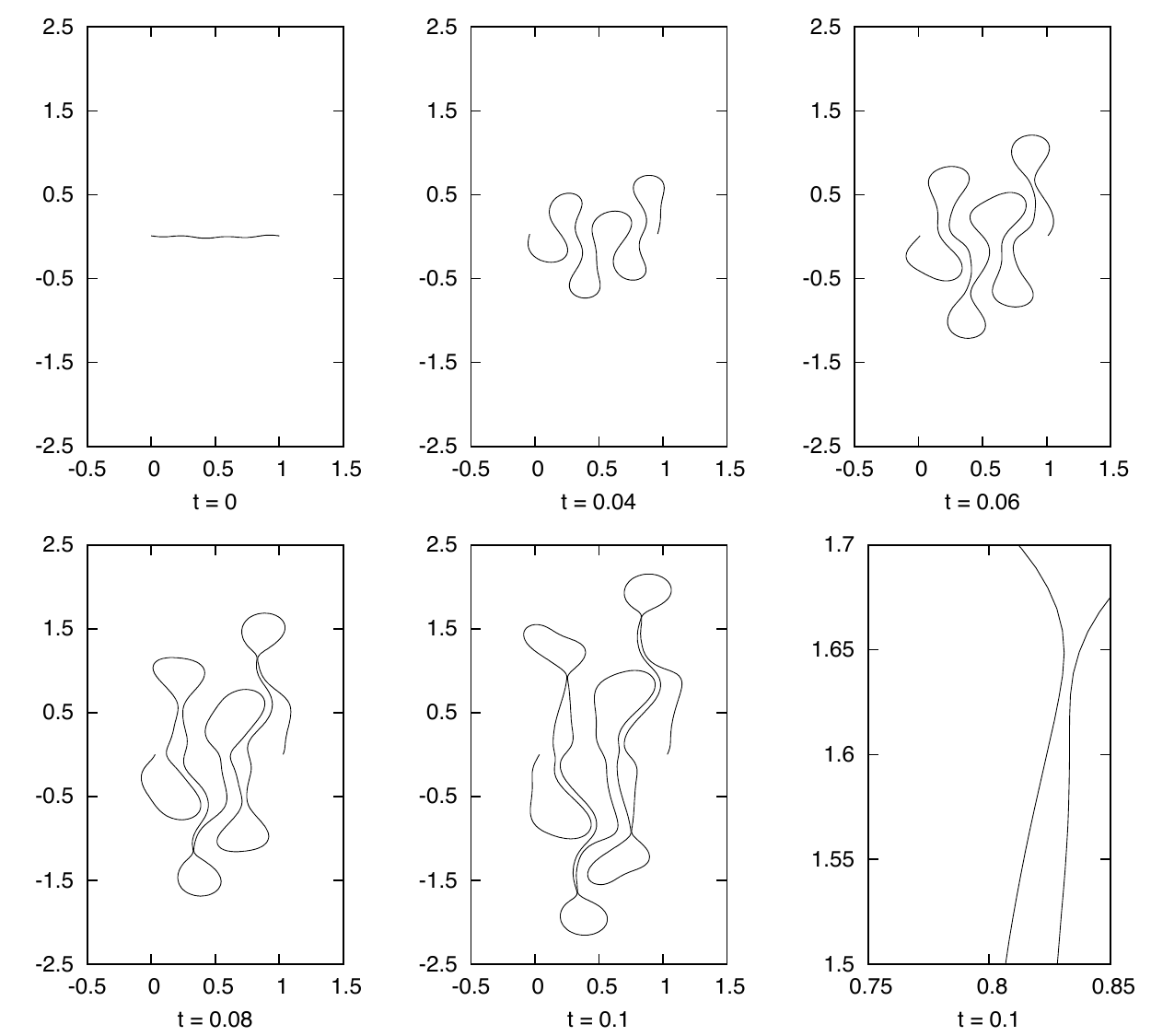}} 
\subfigure[]
          {\includegraphics[width=0.49\linewidth]{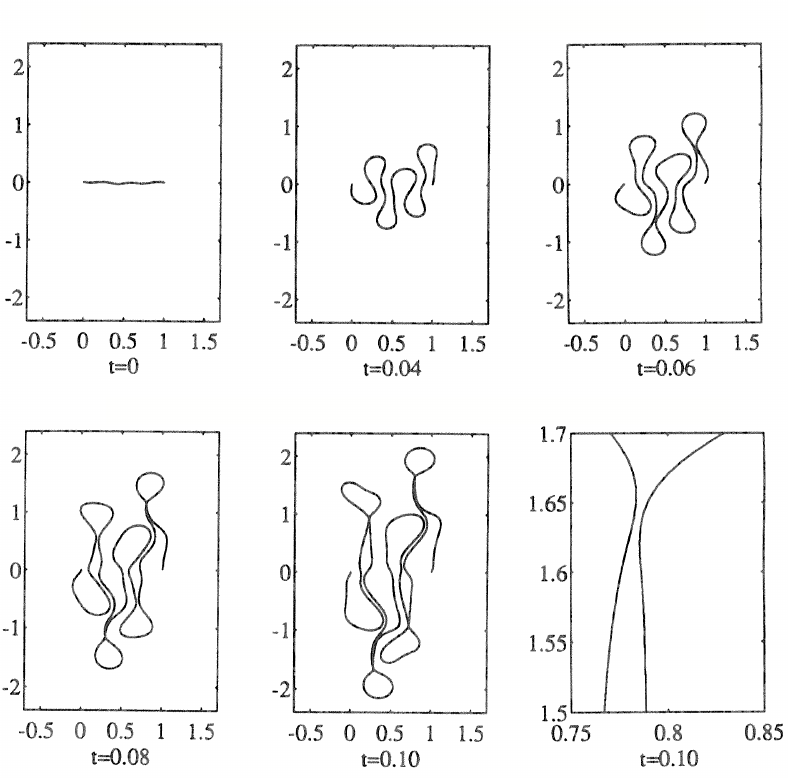}}
	 \par
\end{centering}
\caption{
The instability of a Hele-Shaw interface with initial conditions 
(\ref{HS_init}), evolving according to (\ref{advection}), 
(\ref{BR}),(\ref{gamma}), with parameters $S=0.1$, $R=-50$. 
In our computation ((a), left six panels), we have used $N=2048$ points,
and $\delta t=3.125 \times 10^{-5}$. The constant $\lambda$ was 
chosen according to $\lambda=0.35 \; S (2\pi / N \delta s_{min})^3$, 
which satisfies the stability constraint (\ref{HS_stable_gen}).
For comparison, we show the results of the original computation
\cite{HLS94} ((b), right six panels), obtained for the same physical 
parameters, and using the same number of grid points and time step. 
Differences between the two calculations are visible in the closeup 
of the last panel only. 
\label{fig:evol1D}}
\end{figure}

In Fig.~\ref{fig:evol1D}, we compare our results to the long time run
of Hele-Shaw dynamics, presented as a bench mark for the methods
developed in \cite{HLS94}. Our computations are shown on left at the 
times indicated, those of \cite{HLS94} are shown on the right at
identical times. We have chosen the same physical parameters,
as well as the same spatial resolution ($N = 2048$), and time step
$\delta t=3.125 \times 10^{-5}$. Periodic boundary conditions apply 
in the $x$-direction. No filtering was applied to our data, and no sign
of instability could be observed throughout the highly non-linear
evolution of the interface. As a consequence of the interplay between 
gravitational instability and surface tension, long wavelength 
perturbations are amplified first. Subsequently, the interface 
deforms into a highly contorted shape consisting of long necks 
bounded by rounded fluid blobs. In several places, and as highlighted
in the last panel, fluid necks come close to pinch-off, and small
scale structure is generated. 

\begin{figure}[hbt]
\begin{centering}
\subfigure[]
         {\includegraphics[width=0.39\linewidth]{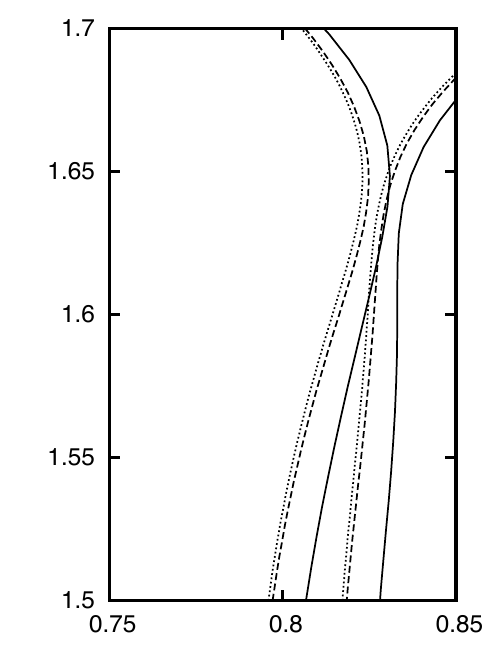}} 
\subfigure[]
          {\includegraphics[width=0.39\linewidth]{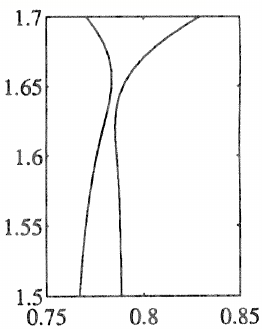}}
	 \par
\end{centering}
\caption{
Comparison between our result (a) and the result presented in 
\cite{HLS94} (b), for the bottom right panel in Fig.~\ref{fig:evol1D}.
The solid curves correspond to $N=2048$ grid points, 
the dashed curve to $N=4096$ and the dotted curve to $N=8192$.
}
\label{fig:zoom}
\end{figure}

The results of the two computations are indistinguishable, except for 
the last panel, in which a closeup is shown. To investigate the 
source of the remaining discrepancy, we have repeated our computation
at twice and four times the original spatial resolution, the results 
of which are shown in the left panel of Fig.~\ref{fig:zoom}. The right 
panel shows the original computation by
\cite{HLS94}, with $N = 2048$. It is seen that to achieve convergence 
on the scale of the closeup, about $N=4096$ grid points are needed,
which yields a result close to that for $N=8192$. Taking the 
highest resolution result as a reference, it is seen that our 
numerical scheme performs at least as well as the original scheme 
of \cite{HLS94}.

\section{Application to Kuramoto-Sivashinsky equation}

The final test of our method treats the Kuramoto--Sivashinsky 
equation \cite{cross93}, which contains fourth-order derivatives: 
\beq
\frac{\partial u}{\partial t} = -u \frac{\partial u}{\partial x} - 
\frac{\partial^2 u}{\partial x^2} - \frac{\partial^4 u}{\partial x^4},
\label{ks}
\eeq
where all coefficients have been normalized to unity. The second-order 
term acts as an energy source and has a destabilizing effect, 
the nonlinear term transfers energy from low to high wavenumbers, while
the fourth-order term removes the energy on small scales. The 
Kuramoto--Sivashinsky equation is known to exhibit spatio-temporal 
chaos, so the main interest lies in predicting the statistical 
properties of solutions. An 
accurate method for solving (\ref{ks}) is described in \cite{KT05}, 
where it is solved on the periodic domain $x \in [0,32\pi]$, with
the initial condition~: 
\beq
u(x,t=0) = \cos\left(\frac{x}{16}\right) 
\left(1+\sin\left(\frac{x}{16}\right)\right).
\label{ks_init}
\eeq

We want to use (\ref{ks}) to illustrate the flexibility of our
method, using a {\it lower order term} to stabilize the algorithm. 
Namely, we use $\lambda u_{xx}$, with $\lambda$ chosen in order to 
counteract the effect of $-u_{xxxx}$. We show that while 
this is certainly not the method of choice to solve this equation, 
it is sufficiently accurate to represent the statistics of the solution. 

\subsection{Numerical scheme}

Equation (\ref{ks}) is discretized on a regular grid, 
using centered finite differences~:
\beqa
\frac{u_j^{n+1}-u_j^n}{\delta t} = -u_j^n \frac{u_{j+1}^n-u_{j-1}^n}{2\delta x} 
- \frac{u_{j-1}^n-2u_{j}^n+u_{j+1}^n}{\delta x^2}
- \frac{u_{j-2}^n-4 u_{j-1}^n+6u_{j}^n-4u_{j+1}^n+u_{j+2}^n}{\delta x^4} &&\nonumber\\
- \lambda \frac{u_{j-1}^n-2u_{j}^n+u_{j+1}^n}{\delta x^2}
+ \lambda \frac{u_{j-1}^{n+1}-2u_{j}^{n+1}+u_{j+1}^{n+1}}{\delta x^2}, &&
\label{scheme_ks}
\eeqa
where $\lambda$ has to be chosen such that the method is stable. 

\subsection{Von Neumann stability analysis and numerical results}

In order to find the right value of $\lambda$ for the scheme to be stable, 
we only need to consider the fourth-order derivative in the equation, 
which is the stiff term to be stabilized. 
As before, inserting $u_j^n = \xi^n e^{ik j \delta x}$ into 
(\ref{scheme_ks}), and retaining only $-u_{xxxx}$ from the 
Kuramoto--Sivashinsky equation, we obtain after simplification~:
\beq
\frac{\xi-1}{\delta t} = 
- \frac{2}{\delta x^4} \left( \cos(2 k \delta x) -4 \cos(k \delta x) + 
3 \right) + \frac{2 \lambda}{\delta x^2} (\xi -1) 
\left( \cos(k \delta x) -1 \right). 
\label{amp_ks}
\eeq
Again, we can identify coefficients $a$ and $b$ from equation (\ref{It}) 
and obtain~:
\beq
a = \frac{2}{\delta x^4} \left( \cos(2 k \delta x) -4 \cos(k \delta x) + 
3 \right) \quad \mathrm{and} \quad 
b = \frac{2\lambda}{\delta x^2} \left( 1 - \cos(k \delta x) \right),
\eeq
so that unconditional stability is guaranteed if
\[
\lambda > \frac{2}{3\delta x^2} 
\frac{\cos(2 k \delta x) -4 \cos(k \delta x) + 3}
{1 - \cos(k \delta x)}.
\]

The maximum value of the right hand side occurs for the largest
wave number $k_{max}=\pi/\delta x$, and the stability constraint 
becomes 
\beq
\lambda > \frac{8}{3\delta x^2}.
\label{stab_ks}
\eeq
We have chosen $\lambda=3/\delta x^2$ for our computations, so 
stability is assured regardless of the time step. However, according 
to the analysis of subsection \ref{sub:damping}, the fact that
we are using a lower order operator for stabilization leads
to a larger time truncation error than a fully implicit second 
order method would have. If we use (\ref{constraint}) for an
estimate of the required time step, we obtain 
\beq
\delta t \simeq \frac{3 \delta x^4}{8} 
\left(\frac{\Delta}{\delta x} \right)^2 .
\label{simple_estimate}
\eeq
\begin{figure}[h!]
\begin{center}
\subfigure[]
         {\includegraphics[width=0.49\linewidth]{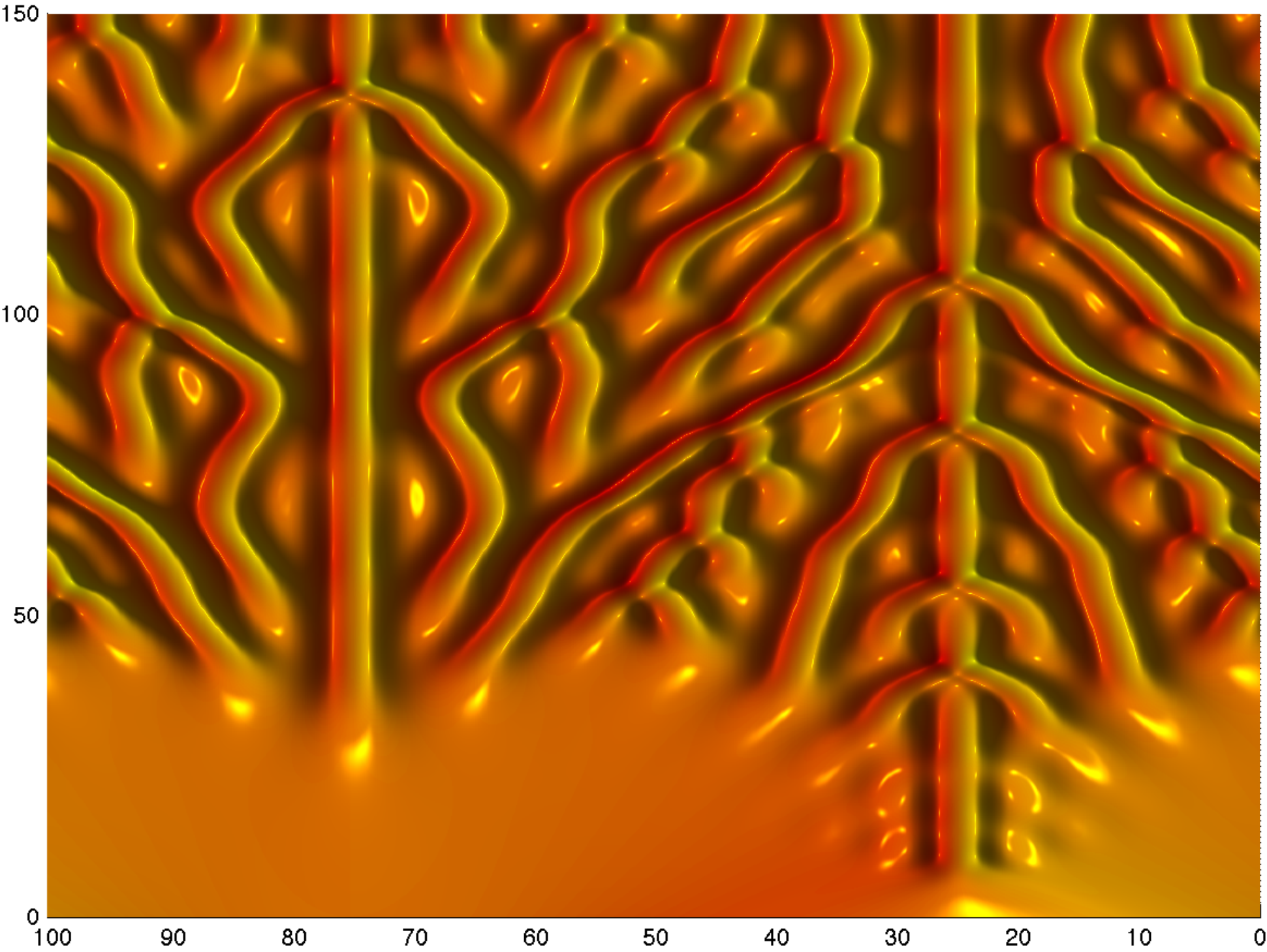}} 
	\subfigure[]
         {\includegraphics[width=0.49\linewidth]{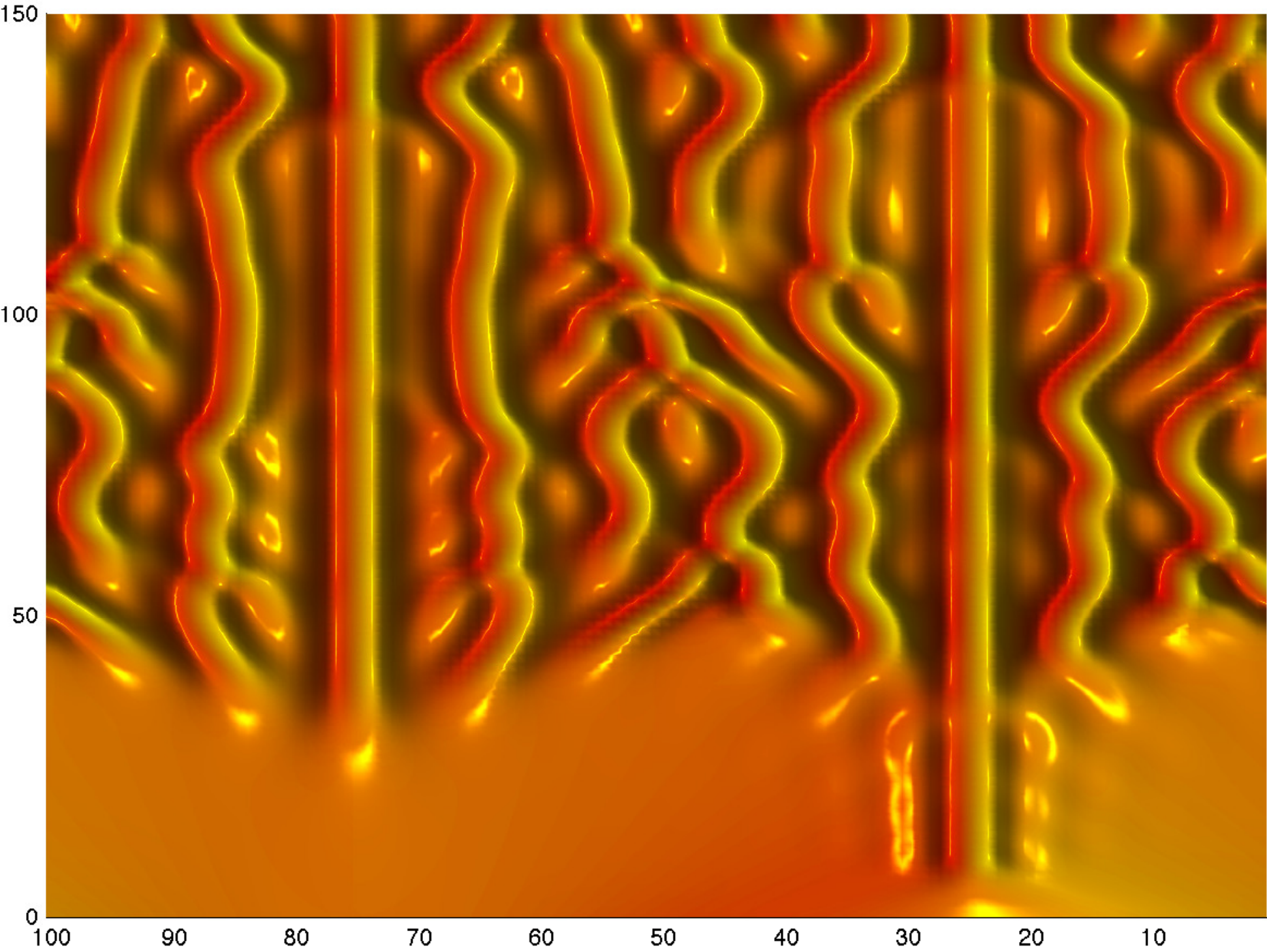}} 
	 \par
\end{center}
\caption{Solution of (\ref{ks}) with initial conditions 
(\ref{ks_init}). The horizontal axis represents the space variable, 
and the vertical axis time. On the left we show our calculation 
with $N=512$ grid points and $\delta t=0.014$, with 
$\lambda=3/\delta x^2$. For comparison, we show the computation
of \cite{KT05} on the right, which uses $128$ grid points and 
a time step $\delta t=1/4$. }
\label{fig:ks}
\end{figure}

In Fig.~\ref{fig:ks} we present a comparison of our computation (left)
with the results of the high-resolution code given in \cite{KT05} (right) 
as a reference. This code is of fourth order in both space and time, 
and we have confirmed that for $N=128$ and $\delta t = 1/4$, the 
solution is represented accurately over the entire time interval shown
in the figure. Since our code is only of second order in space, we 
have chosen $N=512$, which gives a spatial resolution 
of $\delta x \simeq 0.196$. Estimating the smallest relevant 
physical scale as $\Delta \simeq 1 \simeq 5 \delta x$, 
(\ref{simple_estimate}) yields $\delta t = 1.4 \times 10^{-2}$ as 
the time step. Note that this is 75 times larger than the 
minimum explicit time step $\delta t_E = \delta x^4/8$ required
to stabilize the fourth-order operator. 

We have used $\delta t = 1.4 \times 10^{-2}$ to produce Fig.~\ref{fig:ks} 
(left). Although the two solutions eventually evolve differently, it 
appears that their essential features are quite similar. It is important 
to reiterate that our purpose is not to compete with the fourth-order 
scheme of \cite{KT05}, but rather to demonstrate that we can stabilize a 
fourth-order PDE with a second-order operator, without destroying the 
statistical properties of the solution. 


\section{Conclusions}
In this paper, we take a new look at the problem of stiffness,
which leads to numerical instability in many equations of 
interest in physics. It is well established that a PDE 
can be split into several parts, some of which are treated 
explicitly, while only the stiff part is treated implicitly \cite{ARW95}.
However, the realization of such a split may require great
ingenuity \cite{HLS94}, and has to be performed on a case-by-case 
basis. Moreover, the resulting implicit calculation may still 
require elaborate techniques. 

We demonstrate a way around this problem by showing that 
any explicit algorithm can be stabilized using expressions foreign 
to the original equation. This implies a huge freedom in choosing
a term which is both conceptually simple and inexpensive to invert numerically. 
In particular, the stabilizing does not need to represent a 
differential operator, nor does it need to have a physical meaning. 
Since the stiffness comes from short-wavelength modes on the 
scale of the numerical grid, we only require the stabilizing part 
to approximate the true operator in the short wavelength limit. 

We note that although in this paper we were concerned mostly 
with uniform grids, this is by no means necessary, as stability 
criteria such as (\ref{stab_mc}) or (\ref{HS_stable_gen}) are local. 
If the grid spacing varies, this can be accounted for by allowing $\lambda$
to vary in space as well as in time. A possibility we have not explored 
yet is to choose $\lambda$ adaptively. At the moment, the right choice of 
$\lambda$ requires some analysis of the high wavenumber behavior
of the equation. An appropriate algorithm might be able to adjust 
to the optimal value of $\lambda$ automatically, which in general
will be spatially non-uniform. Finally, the feasibility of our
scheme in two space dimensions has already been demonstrated \cite{DD71}. 

\appendix
\section{Choice of tangential velocity}
The tangential velocity of the interface is chosen such that the ratio 
of the distance between two successive points to the total length of 
the interface is conserved in time~:
\beq
s_\alpha(\alpha,t) = R(\alpha)L(t) = 
R(\alpha) \int_0^{2\pi}{s_{\alpha'}d\alpha'},
\label{sa}
\eeq
where $\alpha$ is a marker label, $s(\alpha,t)$ is the arclength, 
$L(t)$ is the total length of the interface and $R(\alpha)$ is such that~:
$$
\int_0^{2\pi}{R(\alpha)}d\alpha=1.
$$
We choose the tangential velocity such that $R(\alpha)$ does not 
change in time. Taking the time derivative of 
$$
s_\alpha = \sqrt{x_\alpha^2+y_\alpha^2}
$$
and using the advection equation (\ref{advection}), one 
finds that 
\beq
s_{\alpha t} = T_\alpha - \theta_\alpha U,
\label{sat}
\eeq
where $\theta$ is the angle between the local tangent and the $x$ axis.

Integrating equation (\ref{sat}) over $\alpha$ and using the time-derivative 
of equation (\ref{sa}) together with the fact that 
${\displaystyle \int_0^{2\pi}{T_{\alpha'}d\alpha'}=0}$, one finally obtains~:
\beq
T(\alpha,t)=T(0,t) + \int_0^\alpha{\theta_{\alpha'} U d\alpha'} 
- \int_0^\alpha{R(\alpha')d\alpha'} \int_0^{2\pi}{\theta_{\alpha'}Ud\alpha'}.
\label{tan_vel}
\eeq

\bibliographystyle{model1-num-names}
\bibliography{../../all_ref.bib}

\end{document}